\documentclass[letter]{aa}

\usepackage{graphicx}
\usepackage{txfonts}
\usepackage{natbib}

\begin{document}

\title{Extended Red Emission and the evolution of carbonaceous nanograins in NGC 7023.
\thanks{This work is based on observations made with the Spitzer Space
Telescope, which is operated by the Jet Propulsion Laboratory, California Institute of Technology 
under a contract with NASA.Based on observations made with the NASA/ESA Hubble Space Telescope, 
obtained from the Data Archive at the Space Telescope Science Institute, which is operated by 
the Association of Universities for Research in Astronomy, Inc., under NASA contract NAS 5-26555. 
These observations are associated with program 9471.
Based on observations obtained at the Canada-France-Hawaii Telescope (CFHT) which is operated by the
National Research Council of Canada, the Institut National des Sciences de l'Univers of the Centre National 
de la Recherche Scientifique of France,  and the University of Hawaii.
Based on observations with ISO, an ESA project with instruments funded by ESA Member States (especially the 
PI countries: France, Germany, the Netherlands and the United Kingdom) and with the participation of ISAS and NASA.}}

\author{O. Bern\'e \inst{1}
\and 
C. Joblin\inst{1}
\and
M. Rapacioli\inst{2}
\and 
J. Thomas\inst{3}
\and
J.-C. Cuillandre\inst{4}
\and
Y. Deville\inst{3}}

\offprints{\\ O.~Berne, \email{olivier.berne@cesr.fr}}
\institute{
Centre d'Etude Spatiale des Rayonnements, Universit\'e Paul 
Sabatier Toulouse~3 et CNRS, Observatoire Midi-Pyr\'en\'ees, 9 Av. du Colonel Roche, 
31028 Toulouse cedex 04, France
\and
Laboratoire de Chimie et Physique Quantique, IRSAMC, Universit\'e Paul 
Sabatier Toulouse~3 et CNRS, 118 Route de Narbonne, 31062 Toulouse Cedex, France
\and
Laboratoire d'Astrophysique de Toulouse-Tarbes, Universit\'e Paul 
Sabatier Toulouse~3 et CNRS, Observatoire Midi-Pyr\'en\'ees, 14 Av. Edouard Belin,
31400 Toulouse, France
\and
Canada-France-Hawaii Telescope Corporation 65-1238 Mamalahoa Highway Kamuela, Hawaii 96743, USA}

\date{Received ?; accepted ?}

\abstract
{Extended Red Emission (ERE) was recently attributed to the photo-luminescence of either doubly ionized 
Polycyclic Aromatic Hydrocarbons (PAH$^{++}$), or charged PAH dimers ($[$PAH$_{2}$]$^+$).}
{We analysed the visible and mid-infrared (mid-IR) dust emission in the North-West and South photo-dissociation regions of 
the reflection nebula NGC 7023.}
{Using a blind signal separation method, we extracted the map of ERE from images
obtained with the Hubble Space Telescope, and at the Canada France
Hawaii Telescope. 
We compared the extracted ERE image to the distribution maps of the mid-IR emission of Very Small Grains (VSGs), 
neutral and ionized PAHs (PAH$^0$ and PAH$^+$) obtained with the Spitzer Space Telescope 
and the Infrared Space Observatory.}
{ERE is dominant in transition regions where VSGs are being photo-evaporated to form free PAH
molecules, and is not observed in regions dominated by PAH$^+$. Its carrier makes a minor contribution
to the mid-IR emission spectrum.}
{These results suggest that the ERE carrier is a transition species formed during the destruction of VSGs.
$[$PAH$_{2}$]$^+$ appear as good candidates but PAH$^{++}$ molecules seem to be excluded.}

\keywords{astrochemistry  \textemdash{} ISM: dust 
\textemdash{} ISM: lines and bands \textemdash{} reflection nebulae \textemdash{} infrared: ISM
\textemdash{} methods: numerical \textemdash{} methods: observational}
\maketitle

\authorrunning{Bern\'e et al.}
\titlerunning{ERE in NGC 7023}

\section{Introduction} \label{int}

Unveiling the composition, structure and charge state of the smallest interstellar dust particles
remains one of today's challenges in astrochemistry.
Progress in this field requires a detailed analysis of the spectral signatures of these dust populations. 
Amongst these signatures is the Extended Red Emission (ERE), a broad emission feature 
ranging from 540 to beyond 900 nm and with a peak wavelength longward of 600 nm to beyond 800 nm 
\citep{smi02b}. ERE is likely due to the photo-luminescence of carbonaceous 
macromolecules or nanograins exposed to UV photons (see Witt et al. 2006 and references therein). 
The ERE is observed in many environments exposed to UV photons (Photo-dissociation Regions; PDR)
found for instance in the diffuse interstellar medium, reflection nebulae and planetary nebulae. It has been observed in
galaxies, NGC 3034 \citep{per95} and NGC 4826 by \citep{pie02}. 

The possible link with the carriers of the Aromatic Infrared Bands (AIBs; also called unidentified infrared
bands at 3.3, 6.2, 7.7, 8.6 and 11.3 \,$\mu$m) has been discussed by several authors.
Both types of  emission features, ERE and AIBs, were found to be relatively cospatial although 
not matching \citep{fur90}, pointing to different but related materials for their carriers \citep{fur92}.
Recently, a detailed study of the spatial distribution of the ERE in the northern PDR
of NGC\,7023 has been performed by \citet{wit06} who concluded that the ERE mechanism is a two-step  process involving
the formation of the carrier and then the excitation of the luminescence,
and proposed doubly-ionized Polycyclic Aromatic Hydrocarbons (PAHs) as plausible carriers. 
On the other hand, recent quantum chemistry
calculations point to PAH dimers ($[$PAH$_{2}$]$^+$) as the carrier of ERE \citep{rhe07}.

In this letter, we provide an efficient and non biased way to extract the ERE map in NGC 7023
from the Hubble Space Telescope (HST) data and new Canada France Hawaii Telescope (CFHT) images using a blind signal separation method. 
We then compare the extracted ERE map to the maps of the different mid-IR
emission carriers, namely Very Small Grains (VSGs), neutral and ionized PAHs
(PAH$^0$ and PAH$^+$), as extracted by  \citet{rap05} and \citet{ber07} from the Infrared Space Observatory (ISO) and
Spitzer Space Telescope (\emph{Spitzer}) observations.

\begin{figure}
\begin{center}
\includegraphics[width=\hsize]{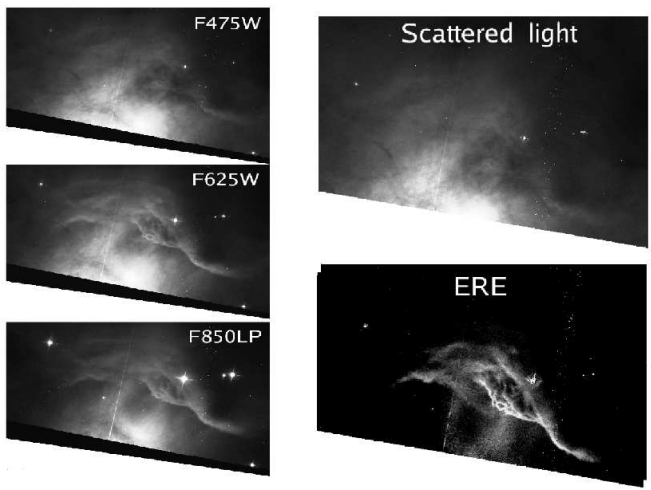}
\vspace{-0.0cm}
\caption{On the left: HST images of the NGC 7023 North-West PDR in three SDSS wide-band filters
(cf. Witt et al. 2006).
On the right: scattered light and ERE images extracted with \emph{FastICA} from the observations}
\end{center}
\label{ext}
\vspace{-0.5cm}
\end{figure}

\section{Observations} \label{obs}

\begin{figure}
\begin{center}
\includegraphics[width=\hsize]{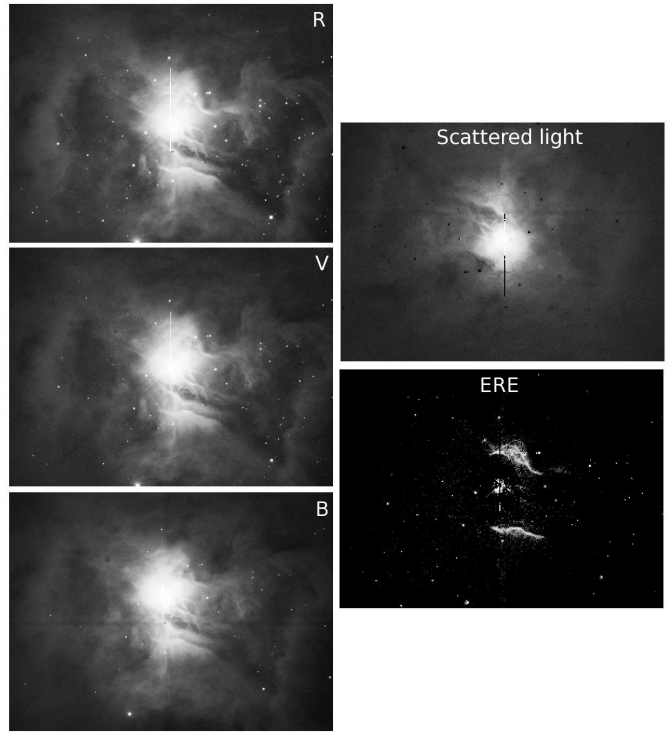}
\vspace{-0.0cm}
\caption{On the left: CFHT images of the NGC 7023 North-West PDR in three BVR filters.
On the right: scattered light and ERE images extracted with \emph{FastICA} from the observations}
\end{center}
\label{cfh}
\end{figure}

\subsection{Infrared observations} \label{inf}
The analysis of the mid-IR emission of the North-West (NW) and South PDRs was first performed by
\citet{rap05} using ISOCAM-CVF data from ISO.
More recent data on the NW PDR obtained with the Infrared Spectrograph (IRS) onboard
\emph{Spitzer} in mapping mode was analysed by \citet{ber07}. The achieved angular resolution is 
3.6" for IRS and 5" for ISOCAM.

\subsection{Visible observations} \label{hub}
ERE was observed in NGC 7023 by \citet{wit90} and \citet{wit06}.
The NW PDR was observed with the HST by \citet{wit06}, using the Advanced Camera for Surveys
(ACS) and the Near Infrared Camera and Multi-Object Spectrometer (NICMOS). We retrieved from the archive
the calibrated, geometrically corrected, dither-combined ACS images (Fig. 1) in three 
wide-band Sloan Digital Sky Survey filters \citep{smi02}: \emph{g}, \emph{r},
and \emph{z} (respectively called F475W, F625W and F850LP on HST). 
The whole NGC 7023 nebula was observed in August 2002 in the B,V 
and R filters at the CFHT using 
the CFH12K CCD mosaic \citep{cui01} as part of the CFHT outreach program. 
A set of 5 dithered exposures of 60 seconds each were obtained in each filter in order to 
remove the mosaic gaps and other physical blemishes. No sky subtraction, 
nor convolution of any sort was applied to the data during the detrending 
and stacking process, ensuring no alteration of the intrinsic brightness 
features of the nebula.

\section{Extraction of the ERE map} \label{ext}
One of the difficulties in properly extracting the ERE map is to remove
the contribution of the scattered light from the central Herbig Be star, HD 200775.
To perform this analysis in an unbiased way,
we applied a Blind Signal Separation (BSS) method successively to the ACS and CFHT images obtained in the three 
wide-band filters (F475W, F625W, F850LP, for ACS and B, V, R for CFHT: see Figs. 1-2).
For a wide-band filter centered at $\lambda$, it is assumed that the observed image $Im_{\lambda}$ can be written as:

\begin{equation}
\label{general}
Im_{\lambda}=a_{\lambda}ERE+b_{\lambda} \emph{Scattered}.
\end{equation}

{In this equation, it is assumed that the spatial patterns, \emph{ERE} and \emph{Scattered}
are unknown but do not depend on wavelength. The "mixing" coefficients $a_{\lambda}$ and $b_{\lambda}$
are unknown and vary with the filter central wavelength $\lambda$.
ERE arises from the UV-illuminated layers of the cloud and therefore does not suffer from
significant extinction in the studied reflection nebula. It is thus reasonable to assume that its spatial pattern (\emph{ERE} in Eq. 1) is 
independent of wavelength. In the case of scattered light, this
might not be true if strong radiative transfer is present as shown for instance 
by \citet{nuz00} in the case of galaxies. However, in the present object, the scattered
light is subject to much less extinction than in galaxies, we therefore assume that 
its spatial pattern (\emph{Scattered} in Eq. 1) is independent of wavelength.
The BSS method we used here, called FastICA \citep{hyv99}, enables us to recover 
the spatial patterns of ERE and scattered light using
the statistical information available in the three images.}
FastICA uses the assumptions of independence and 
non-Gaussianity of the original sources (here the spatial patterns of ERE and scattered light) 
and thus belongs to the class of methods called Independent Component Analysis (ICA). 
The FastICA algorithm maximizes the non-Gaussianity of the output signals 
(i.e. estimations of the sources) by using a fixed-point algorithm. 
The extracted images of ERE and scattered light from ACS and CFHT images are 
presented on the right of Figs. 1 and 2 respectively.
{We can qualitatively compare these extracted ERE images to the one obtained by
\citet{wit06}. The images are very similar but we find a small contribution
of ERE in the cavity. 
The CFHT and HST ERE patterns appear to be consistent with each other.}

\section{Linking ERE and mid-IR emission carriers} \label{com}

\subsection{Comparing the spatial distribution of mid-IR emitting populations and ERE}

\citet{rap05} and \citet{ber07} have used signal processing methods to analyse 
the mid-IR spectral cubes form ISO and \emph{Spitzer} and extract the spectra and associated spatial distributions for the
different emitting populations, VSGs, PAH$^0$ and PAH$^+$ as shown in Figs. 3-4.
In Fig. 3,  the HST ERE map was overlayed on the \emph{Spitzer} maps showing that ERE
arises from the region where the population identified as PAH$^0$ dominates the mid-IR
emission. In Fig. 4, we overlayed the ERE extracted from the CFHT observations on the maps
extracted from ISOCAM data. The same correlation is found for the NW PDR as with 
\emph{Spitzer}/HST, though the level of detail reached here is lower due to the lower
spatial resolution of both ISOCAM vs IRS and CFHT vs HST. However, Fig. 4 provides
the information on the South PDR, where again, the ERE filament is found in a region
dominated by PAH$^0$, and close to the frontier with VSGs.

\begin{figure}
\includegraphics[width=\hsize]{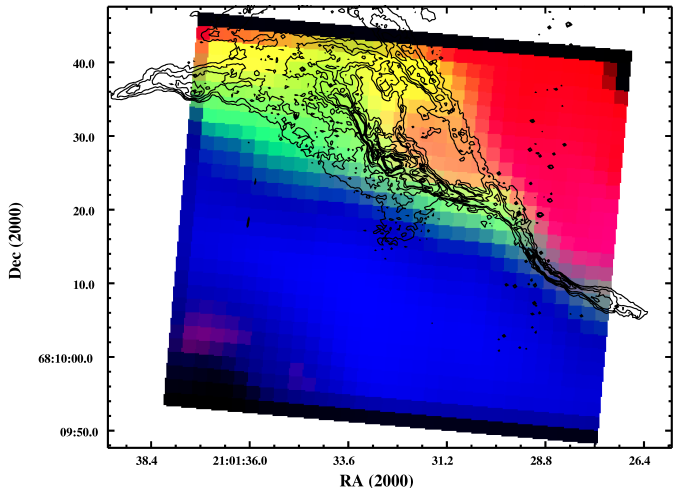}
\vspace{-0.5cm}
\caption{Distribution maps of the three populations of mid-IR emitters in NGC 7023 NW from
Spitzer-IRS observations: VSGs in red, PAH$^0$ in green and PAH$^+$ in blue (cf. Bern\'e et al 2007).
{Overlayed in contours is the emission map of ERE extracted from HST images (Fig. 1).}}
\label{rvb}
\vspace{-0.0cm}
\end{figure}

\begin{figure}
\includegraphics[width=\hsize]{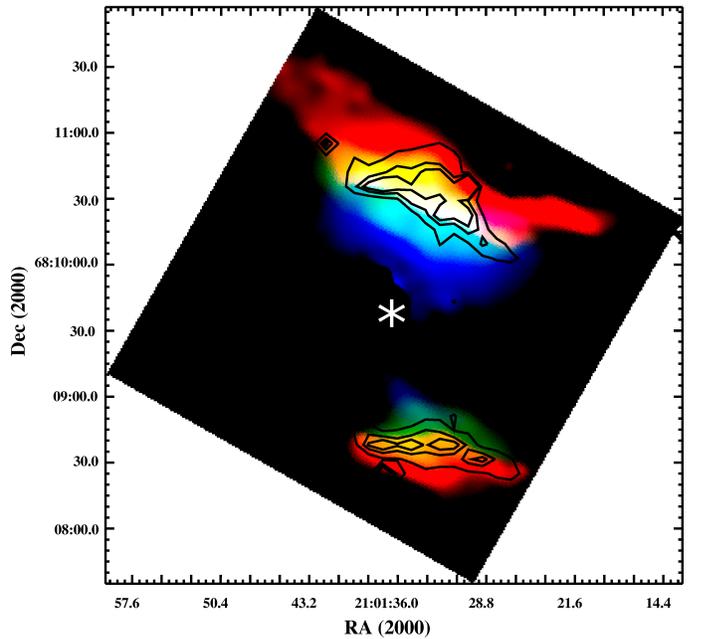}
\vspace{-0.0cm}
\caption{Distribution maps of the three populations of mid-IR emitters in NGC 7023 from
ISOCAM observations: VSGs in red, PAH$^0$ in green and PAH$^+$ in blue first presented by Rapacioli et al. (2005)
and reanalysed in this work. Overlayed in contours is the emission map of ERE extracted from CFHT images (Fig. 2).}
\label{rvb}
\end{figure}

\subsection{Is the ERE carrier an important contributor to the mid-IR emission spectrum ?} \label{com}

A simple energy budget including ERE and the mid-IR emission can be made.
First, we can consider that the ERE carrier is excited by photons of at least 7\,eV energy. This corresponds to
the mean energy absorbed by VSGs according to \citet{rap05}.
If we assume that one ERE photon is emitted at an energy of about 1.7 eV (corresponding to a central wavelength of the 
ERE band at 680 nm detected by \citealt{wit90}), then the rest of the absorbed energy (5.3 eV) will be
emitted in the IR. Thus, about 25 $\%$ of the absorbed light is converted into ERE. Table 1 summarizes
the values of the mid-IR and ERE fluxes  for the regions of 
NGC 7023 where ERE is detected as well as for the prototypical case of the Red Rectangle.
The gross ratio between ERE and mid-IR flux for NGC 7023 NW PDR is below 2 $\%$ (0.4$\%$ for the South PDR). 
Considering the above approximation for the ratio of energy emitted in the IR vs visible for one particle, this yields a proportion 
of less than about  $3 \times 2 \%= 6\%$ of the mid-IR emission due to the ERE carrier in the NW PDR (1.2$\%$ for the South PDR ).
As a comparison, in the Red Rectangle protoplanetary nebula, the strongest known source of ERE, we find that this ratio
is around $3 \times 0.8 \%= 2.4\%$. This implies that the ERE carrier contributes to only a few percent of the mid-IR 
emission, and thus its signature will be difficult to identify in this spectral region. Following our assignment, PAH$^0$ are excluded
as possible candidates for the ERE, though they were initially proposed as potential carriers by \citet{den86}.

\begin{table}[ht!]
\caption{Mid-IR and ERE fluxes in NGC 7023 and the Red Rectangle.}
\label{table1}
\begin{center}
\begin{tabular}{cccc}

\hline \hline  
                    & Mid-IR$^*$            &  ERE$^{**}$ & $I_{ERE}/I_{mid-IR}$\\
   PDR          &  (Wm$^{-2}$sr$^{-1}$)   & (Wm$^{-2}$sr$^{-1}$)& \\

\hline
\noalign{\smallskip}

NGC 7023-S        & $4.11\,10^{-5}$ & $1.44\,10^{-7}$ & 0.0035\\
NGC 7023-N      & $1.40\, 10^{-4}$   & $2.30\, 10^{-6}$ & 0.0164\\
Red-Rect    & $9.04\, 10^{-4}$   & $7.10\,10^{-6}$ & 0.008\\

\hline
\multicolumn{4}{p{8cm}}{
$^*$ Integrated Spitzer-IRS (5-14 $\mu$m) spectrum at the positions
at which ERE was observed by Witt and Boroson (1990), 
$^{**}$ from Witt and Boroson (1990), integrated between 550 and 850 nm}
\\
\end{tabular}
\end{center}  
\end{table}

\section{Possible carriers of ERE in NGC 7023} \label{pos}

\subsection{The case of PAH$^{++}$}
Using the ACS observations, \citet{wit06} showed that ERE is likely a two step process involving the
formation of the carrier and then the excitation of the luminescence. The first step requires far-UV photons 
(E~$>$~10.5 eV) and supports the idea that the carrier of ERE is produced by the 
photo-dissociation/photo-ionization of a precursor. From this result, they proposed PAH$^{++}$ as the
carrier of ERE, invoking that these species have an ionisation potential above 10.5 eV and
have strong absorption bands in the optical and near-UV regions.
In the previous section, we have shown that the ERE in NGC 7023 arises from the region where PAH$^0$
are abundant, which differs from the region where  PAH$^+$ are abundant. Thus, in the framework of 
our previous work (\citealt{rap05}; \citealt{ber07}) this rules out the possibility that ERE is carried by doubly  
ionized PAHs as proposed by \citet{wit06}.

\subsection{The case of $[$PAH$_{2}]^+$}
In a recent theoretical work \citep{rhe07},  it was shown  that charged PAH dimers ([PAH$_{2}]^+$),
more specifically the subclass made of closed-shell species, can fluoresce in the ERE range with a 
quantum yield that is consistent with this emission. 
The overlays of Figs. 3 and 4 clearly show that ERE is dominant in regions where VSGs
are dissociated and PAH$^0$ species are abundant. This suggests that the ERE carrier is a
transient species produced during the evaporation of VSGs. This fits well with the two-step scenario
of \citet{wit06}. Indeed, \citet{rap06} have found that  PAH clusters such as (C$_{24}$H$_{12}$)$_{4}$
and (C$_{24}$H$_{12}$)$_{13}$ start being dissociated into monomers at internal energies of around 
10 eV. This is in agreement with the  threshold of 10.5 eV set by \citet{wit06} for the production of the
ERE carrier. Then, for ERE to be observed, the carrier should survive long 
enough in the PDR i.e. be reformed as efficiently as it is destroyed.
$[$PAH$_{2}$]$^+$ appear as good candidates because (1) their stability is expected to be increased
relative to neutral dimers because 
of charge delocalization effects \citep{bou02}, (2) their abundance is favored because they
constitute the final stage in the photodissociation cascade starting from larger clusters;
(3) they can be reformed efficiently by collision of a neutral and ionized PAH
as this process is favored by the long range ion $\leftrightarrow$ induced dipole interaction
(see discussion in \citealt{rap06}). 
PAH$^{+}$ are not abundant species in the ERE region but are present ($\sim 10 \%$ of total mid-IR emission). 
This fraction of PAH$^{+}$ would be enough to lead to the reformation of $[$PAH$_{2}$]$^+$ 
in regions where PAH$^{0}$ are abundant. 
Thus, $[$PAH$_{2}$]$^+$ are favorable candidates since they are expected to be relatively 
more stable and, perhaps more important, to have an efficient reformation path.

\section{Conclusion} \label{con}
In this letter we have compared in NGC 7023 the spatial distribution of ERE and that of the mid-IR
emitters : VSGs,  PAH$^0$,  PAH$^+$. We find strong evidence that the ERE carrier is produced
in the region of destruction of VSGs and show that it has a negligible contribution to the mid-IR
emission. We show that PAH$^{++}$ are unlikely to be the carrier
of ERE and conclude that  $[$PAH$_{2}$]$^+$ are attractive candidates, on the basis
of qualitative arguments on the balance between photodissociation and reformation of these clusters.
Further investigations are needed, following the strategy presented in this letter, but
in other PDRs. Spectro-imagery of such regions in the mid-IR with a high spatial resolution is 
needed in order to be able to trace the thin frontier between PAHs and VSGs where small clusters
are likely to be present.
Finally, laboratory spectroscopic studies on $[$PAH$_{2}$]$^+$ are needed to progress in this
identification. In particular,  closed-shell cation dimers as proposed by \citet{rhe07}
deserve additional studies.

\begin{acknowledgements}

We acknowledge the anonymous referee for his comments on the manuscript.

\end{acknowledgements}

\bibliographystyle{aa}
\bibliography{OB9158}

\end{document}